\def\rfr#1{eq. (\ref{#1})}

\def\virg#1{``#1''}

\def\eqi{\begin{equation}}
\def\eqf{\end{equation}}
\def\eqia{\begin{eqnarray}}
\def\eqfa{\end{eqnarray}}
\def\rp#1#2{{#1\over#2}} \def\lb#1{\label{#1}}

\def\bds#1{\boldsymbol{#1}}

\documentclass{aastex}
\usepackage{url}
\usepackage{amsmath,amsthm,amscd,amssymb}
\usepackage{latexsym}
\usepackage{graphicx,epsfig}

\RequirePackage{color}

\newcommand{\emaila}{lorenzo.iorio@libero.it}

\begin{document}

\title{Phenomenological constraints on Lema\^{\i}tre-Tolman-Bondi cosmological inhomogeneities from solar system dynamics}
\shortauthors{L. Iorio}

\author{Lorenzo Iorio\altaffilmark{1} }
\affil{F.R.A.S.: Viale Unit\`{a} di Italia 68, 70125, Bari (BA), Italy.}

\email{\emaila}

\begin{abstract}
We, first, analytically work out the long-term, i.e. averaged over one orbital revolution, perturbations on the orbit of a test particle moving in a local Fermi frame induced therein by the cosmological tidal effects of the inhomogeneous Lema\^{\i}tre-Tolman-Bondi (LTB) model. The LTB solution has recently attracted attention, among other things, as a possible explanation of the observed cosmic acceleration without resorting to dark energy.
Then, we phenomenologically constrain both the parameters $K_1\doteq -\ddot{\mathfrak{R}}/{\mathfrak{R}}$ and $K_2\doteq -\ddot{\mathfrak{R}}^{'}/{\mathfrak{R}}^{'}$  of the LTB metric in the Fermi frame by using different kinds of solar system data. The corrections $\Delta\dot\varpi$ to the standard Newtonian/Einsteinian precessions of the perihelia of the inner planets recently estimated with the EPM ephemerides, compared to our predictions for them,  yield $K_1 = (4\pm 8)\times 10^{-26}$ s$^{-2}$, $K_2 = (3\pm 7)\times 10^{-23}$ s$^{-2}$. The residuals of the Cassini-based Earth-Saturn range, compared with the numerically integrated LTB range signature, allow to obtain  $K_1\approx K_2\approx 10^{-27}$ s$^{-2}$. The LTB-induced distortions of the orbit of a typical object of the Oort cloud with respect to the commonly accepted Newtonian picture, based on the observations of the comet showers from that remote region of the solar system, point towards  $K_1\approx K_2\lesssim 10^{-30}-10^{-32}$ s$^{-2}$. Such figures have to be compared with those inferred from cosmological data which are of the order of $K_1\approx K_2=-4\times 10^{-36}$ s$^{-2}$.
 \end{abstract}


\section{Introduction}
Inhomogeneous cosmological models \citep{Hellaby} have recently attracted much attention because, among other things, some of them may potentially be useful in explaining the observed Universe's acceleration without resorting to  dark energy. Thus, it is important to put them on the test independently of the phenomenon itself for which they have been purposely introduced.

In the framework of the general theory of relativity, the most general form of the line element $(ds)^2$ of a spherically symmetric inhomogeneous space-time in which the source in the Einstein's field equations is a perfect fluid is, in\footnote{Recall that, by definition, the comoving coordinates are those in which the vector field $u^{\alpha},\ \alpha=0,1,2,3$ has only the time component, i.e. $u^{\alpha}\propto \delta^{\alpha}_{\ 0}$. If the vector field $u^{\alpha}$ has zero rotation, as it happens if the space-time is spherically symmetric and the metric obeys the Einstein equations with a perfect fluid source, the comoving coordinates can be chosen so that they are synchronous, i.e. in them the metric tensor has no off-diagonal components $g_{0i},\ i=1,2,3$.} comoving-synchronous spherical coordinates $\{t,r,\theta,\phi\}$ \citep{Krasi},
{{\eqi (ds)^2 = e^{{\mathfrak{C}}(r,t)}(cdt)^2-e^{\mathfrak{A}(r,t)}(dr)^2 - {\mathfrak{R}}^2(r,t)[(d\theta)^2+\sin^2\theta (d\phi)^2].\lb{metrica}\eqf}}
In order to solve the resulting Einstein field equations,
 an equation of state has to be assumed. The most natural choice consists of setting the pressure equal to zero, corresponding to a dust evolution driven by gravitation only. From the equations of motion of a perfect fluid it turns out that, for $p=0$, the dust moves along timelike geodesics. As a consequence, $\mathfrak{C}=0$ and, for\footnote{Here and in the following,  $\mathfrak{R}^{'}$ denotes the partial derivative of $\mathfrak{R}$ with respect to $r$.} ${\mathfrak{R}}^{'}\neq 0$,
\eqi e^{\mathfrak{A}}=\rp{   {{\mathfrak{R}}^{'}}^2  }{1+2\mathfrak{E}(r)},\eqf
where $\mathfrak{E}(r)$ is an arbitrary function such that $\mathfrak{E}\geq -1/2$ for all $r$ in order to preserve the right signature of the metric of \rfr{metrica}. The resulting space-time line element is the so-called Lema\^{\i}tre-Tolman-Bondi (LTB) model. Indeed, such spherically symmetric inhomogeneous dust models were first discovered by \citet{Lem} and further studied by \citet{Tol} and \citet{Bon}; for a detailed discussion of several properties see \citet{Krasi}. Such models are among the best known and most
useful exact solutions of Einstein's equations. Since they allow us to examine non-linear effects analytically,
or at least in a tractable way, there is an extensive literature using them mostly to describe cosmological
inhomogeneities \citep{Hella}, but also as in other theoretical contexts, such as
gravitational collapse \citep{colla} and censorship of singularities \citep{Cens} or quantum gravity \citep{quantum}.
Several authors considered LTB models
as tools to probe how the cosmic acceleration associated to recent observations can be
accounted for inhomogeneities, without introducing dark energy \citep{Paran,Sark,Enq,Bellido}.
LTB models are also a standard choice  to apply Buchert's scalar
averaging formalism \citep{avera}, in which the effects of dark energy could be mimicked by
\virg{backreaction} terms. See \citet{Noelle} for a comprehensive review of such aspects.

Here we are interested in putting constraints in a phenomenological way on some parameters of the LTB metric from its local tidal effects on the dynamics of test bodies of the solar system.  To this aim, it is necessary to set up a standard orthonormal tetrad frame parallel transported along the worldline of a fundamental observer representing, in this case, the Sun, and explicitly derive the expression of the LTB metric in such a local Fermi frame. This has been recently done by \citet{Mash07}. In general, the motion of test bodies can be studied in such  Fermi coordinate systems following ideas and methods developed by \citet{Syn,Mana,Mash}.

The paper is organized as follows. In Section \ref{dua} we analytically work out the LTB effects on the orbital motion of a test particle in the local Fermi frame and compare them with some results present in literature. In Section \ref{tria} we compare our theoretical predictions to different types of bodies and data of the solar system to phenomenologically put constraints on \citep{Mash07}
 \begin{eqnarray}
   K_1 &\doteq & -\rp{\ddot{\mathfrak{R}}}{\mathfrak{R}}, \\ \nonumber \\
   K_2 &\doteq & -\rp{\ddot{\mathfrak{R}}^{'}}{\mathfrak{R}^{'}}
 \end{eqnarray}
 entering the LTB metric written in terms of the Fermi coordinates of a fundamental observer.
 Concerning their values infered from cosmological data,  from (62)-(64) by \citet{Mash07}
 it turns out that, in proximity of the origin of the local Fermi frame, i.e. for $r\rightarrow 0$,
 $K_1$ and $K_2$ are rather similar being of the order of\footnote{$K_{1/2}$ means that both $K_1$ and $K_2$ have approximately the same values.} \eqi K_{1/2}\approx q_0 H_0^2.\eqf In it \citep{hubblo}
 \eqi H_0=71.0 \ {\rm km s}^{-1}\ {\rm Mpc}^{-1}=2.3\times 10^{-18}\ {\rm s}^{-1}\eqf is the present value of the Hubble parameter, and \citep{Xu} \eqi q_0 \doteq\rp{1+3\widetilde{\Omega}_{\rm DE}w_{\rm DE}}{2}=-0.7\eqf is the deceleration parameter in which \citep{hubblo} $\widetilde{\Omega}_{\rm DE}=0.734\pm 0.029$ is the dark energy density and \citep{hubblo} $w_{\rm DE}=-1.12^{+0.42}_{-0.43}$ is the equation of state. Thus, the order of magnitude of $K_1$ and $K_2$ inferred from cosmological observations is
 \eqi K_{1/2}\approx -4\times 10^{-36}\ {\rm s}^{-2}.\eqf
 Section \ref{quatra} is devoted to summarizing our findings and to the conclusions.
\section{Analytical calculation}\lb{dua}
In the Fermi coordinates $\{T,X,Y,Z\}$ associated with a fundamental observer corresponding, in this case, to the Sun, the LTB tidal perturbing potential\footnote{Here and in the following application of the Lagrange's equations for the variation of the Keplerian orbital elements $U$ is defined according to the convention $\bds A=\bds\nabla U$. That is why our $U_{\rm LTB}$ of \rfr{Upot} is $\mathcal{V}_{\rm N}$ of (A.1) by \citet{Mash07} with a minus sign. } \citep{Mash07}
\eqi U_{\rm LTB}= -\rp{1}{2}\left[K_1(X^2+Y^2)+K_2 Z^2\right]\lb{Upot}\eqf arises.
It induces a perturbing acceleration \citep{Mash07}
\eqi
\bds A_{\rm LTB}=-K_1\bds R +(K_1-K_2)Z\bds{k}\lb{LTBacce},
\eqf
where $\bds k$ is the unit vector along the $Z$ axis.

In order to work out the long-term variations of  the Keplerian orbital elements, it is convenient to adopt the Lagrange's perturbative approach \citep{Capde}, valid for perturbations arising from a potential function, since, as we will see, it implies just one integration.
In such a framework, $U_{\rm LTB}$ has to be evaluated onto the unperturbed Keplerian ellipse and averaged over one orbital period, i.e one has to compute
\eqi\left\langle U_{\rm LTB}\right\rangle= \rp{1}{P_{\rm b}}\int_0^{P_{\rm b}}\overline{U}_{\rm LTB} dT,\eqf in which $\overline{U}_{\rm LTB}$ denotes the perturbing potential computed in terms of the parameters of the Keplerian orbit.
To this aim, useful relations in terms of the eccentric anomaly $E$ are \citep{Capde}
\begin{eqnarray}
R &=& a(1-e\cos E),\\ \nonumber \\
  \cos f &=& \rp{\cos E-e}{1-e\cos E}, \\ \nonumber \\
  \sin f &=& \rp{\sqrt{1-e^2} \sin E}{1-e\cos E}, \\ \nonumber \\
  dT &=& \rp{(1-e\cos E)}{n}dE,
\end{eqnarray}
where $a$ and $e$ are the semimajor axis and the eccentricity, respectively, of the unperturbed Keplerian ellipse,  $f$ is the true anomaly reckoning the position of the test particle along it, and $n\doteq 2\pi/P_{\rm b}=\sqrt{GM/a^3}$ is the unperturbed Keplerian mean motion.
The Cartesian coordinates of the moving test particle along the unperturbed Keplerian ellipse, explicitly entering \rfr{Upot}, are \citep{Capde}
\begin{eqnarray}
  X &=& R\left(\cos\Omega\cos u\ -\cos I\sin\Omega\sin u\right),\\ \nonumber \\
 Y &=& R\left(\sin\Omega\cos u + \cos I\cos\Omega\sin u\right),\\ \nonumber \\
 Z &=& R\sin I\sin u,
\end{eqnarray}
where $\Omega$ is the longitude of the ascending node, $I$ is the inclination of the orbit to the reference $\{XY\}$ plane, $u\doteq \omega+f$ is the argument of latitude, in which $\omega$ is the argument of pericentre.
Thus, the perturbing potential, averaged over one orbital revolution, turns out to be
\eqi\left\langle U_{\rm LTB}\right\rangle = K_1\mathcal{D}(a,e,I,\omega) +K_2\mathcal{E}(a,e,I,\omega),\lb{Uave}\eqf
with
\eqi \mathcal{D}=-\rp{a^2}{8}\left[3+\cos 2 I +\rp{e^2}{2}\left(9+3\cos 2 I+10\sin^2 I\cos 2\omega\right)\right],\lb{Dpot}\eqf
and
\eqi \mathcal{E}=-\rp{a^2\sin^2 I}{4}\left[1-\rp{e^2}{2} \left(-3+5\cos 2\omega\right)\right].\lb{Epot}\eqf
Now, \rfr{Uave}, with \rfr{Dpot} and \rfr{Epot},  can be plunged into the right-hand-sides of the Lagrange's equations which are  \citep{Capde}
\begin{eqnarray}
  \rp{d a}{dT} &=& \rp{1}{na}\left(2\rp{\partial U}{\partial \mathcal{M}}\right)\lb{smadot}, \\ \nonumber \\
  \rp{d e}{dT} &=& \rp{1}{na^2}\rp{1-e^2}{e}\left( -\rp{1}{\sqrt{1-e^2}}\rp{\partial U}{\partial\omega} +  \rp{\partial U}{\partial\mathcal{M}} \right), \\\nonumber \\
  \rp{d I}{dT} &=& \rp{1}{na^2\sqrt{1-e^2}\sin I}\left( -\rp{\partial U}{\partial\Omega} +\cos I  \rp{\partial U}{\partial\omega} \right), \\\nonumber \\
  \rp{d \Omega}{dT} &=& \rp{1}{na^2\sqrt{1-e^2}\sin I}\left( \rp{\partial U}{\partial I}  \right), \\\nonumber \\
  \rp{d \omega}{dT} &=& \rp{1}{na^2\sqrt{1-e^2}}\left( \rp{1-e^2}{e}\rp{\partial U}{\partial e} -\rp{\cos I}{\sin I}  \rp{\partial U}{\partial I} \right), \\\nonumber \\
  \rp{d \mathcal{M}}{dT} &=& n + \rp{1}{na^2}\left( -2a\rp{\partial U}{\partial a} - \rp{1-e^2}{e} \rp{\partial U}{\partial e} \right),
\end{eqnarray}
where $\mathcal{M}$ is the mean anomaly. In this way, the long-term variations of the Keplerian orbital elements can be straightforwardly obtained from simple derivatives. From \rfr{smadot} and \rfr{Uave}, with \rfr{Dpot} and \rfr{Epot}, it immediately turns out that $a$ remains unchanged, on average.
Instead, the other Keplerian orbital elements undergo long-term changes.
They are
{
\footnotesize{
\begin{eqnarray}
  \left\langle\dot e\right\rangle &=&  -\rp{5 e\sqrt{1-e^2}\Delta K}{4 n}\sin^2 I\sin 2\omega\lb{eccedo},\\ \nonumber \\
  \left\langle\dot I\right\rangle &=& \rp{5 e^2\Delta K}{8n\sqrt{1-e^2}}\sin 2 I\sin 2\omega, \\ \nonumber \\
  \left\langle\dot\Omega\right\rangle &=& \rp{\Delta K\cos I}{2 n\sqrt{1-e^2}}\left[1+\rp{e^2}{2}\left(3-5\cos 2\omega\right)\right], \lb{nodus}\\ \nonumber \\
  \left\langle\dot\omega\right\rangle &=& \rp{(-11+6e^2)K_1  + (-1+6e^2)K_2 +5\Delta K\left[(-1+2e^2) \cos 2\omega -2\cos2 I\sin^2\omega\right]   }{8n\sqrt{1-e^2}}, \lb{perius} \\ \nonumber \\
  \left\langle\dot{\mathcal{M}}\right\rangle &=& n+\rp{(7+3e^2)\left[3K_1+K_2+\Delta K\cos 2 I +10(1+e^2) \Delta K\sin^2 I\cos 2\omega\right]}{8n}\lb{anomdot},
\end{eqnarray}
}
}
where we have defined $\Delta K\doteq K_1-K_2$.
While the eccentricity $e$ and the inclination $I$ experience only long-term, harmonic variations with frequency $2\omega$, the longitude of the ascending node $\Omega$, the argument of pericenter $\omega$ and the mean anomaly $\mathcal{M}$ undergo secular precessions as well.

Much more cumbersome calculations show that the Gauss\footnote{Contrary to the Lagrange scheme, it is not limited to accelerations derivable from a potential function, being valid for any kind of perturbations.} perturbative approach \citep{Capde}, based on the projections of the perturbing acceleration $\bds A_{\rm LTB}$ of \rfr{LTBacce} onto the radial $\bds{\hat{R}}$, transverse $\bds{\hat{\Xi}}$ and normal $\bds{\hat{\Upsilon}}$ directions of a frame co-moving with the test particle and implying six integrals, yields the same results as \rfr{eccedo}-\rfr{anomdot}, and also $\left\langle \dot a \right\rangle=0$.

It turns out from \rfr{nodus}-\rfr{perius} that the precession of the longitude of pericentre $\varpi\doteq\Omega+\omega$ can be cast into the useful form
\eqi\left\langle\dot\varpi\right\rangle=K_1 \mathcal{F}(I,e,\omega)+ K_2 \mathcal{G}(I,e,\omega),\lb{peridot}\eqf
with
\eqi \mathcal{F}\doteq\frac{-\rp{11}{4}+\cos I-\rp{5}{4}\cos 2 I -\rp{5}{2}\sin^2 I\cos 2\omega + e^2\left[\rp{3}{2}(1+\cos I)+5\sin^2\left(\rp{I}{2}\right)\cos 2 \omega\right]  }{2 n\sqrt{1-e^2}}, \lb{effe}\eqf
and
\eqi \mathcal{G}\doteq -\rp{\sin^2\left(\rp{I}{2}\right)}{2 n \sqrt{1-e^2}}\left[3-5\cos 2\omega+10\cos I\sin^2\omega +e^2\left(5\cos 2\omega -3\right)\right].\lb{gi}\eqf
From \rfr{effe} and \rfr{gi} it can be noted that both secular and long-period, harmonic terms are present. For orbits exhibiting moderate eccentricities and inclinations to the reference plane, like those of the planets of the solar system, the harmonic terms play a minor role because they appear multiplied by $\sin^2 I, \sin^2(I/2), e^2\sin^2(I/2)$. Anyway, since we do not know a-priori the magnitudes of $K_1$ and $K_2$, we will retain all the terms of \rfr{effe} and \rfr{gi} in our calculations.

Concerning the results by \citet{Mash07}, they used the Gauss equations for the variation of the elements and the Delauney orbital elements. \citet{Mash07}, interested in slightly eccentric orbits, obtained  non-zero secular variations of the semimajor axis $a$ of order $\mathcal{O}(e^2)$, and of the eccentricity $e$ of order $\mathcal{O}(e)$, although not explicitly shown; actually, \rfr{eccedo} is of order $\mathcal{O}(e)$, while we obtain $\left\langle\dot a\right\rangle=0$ to all orders in $e$ since it is an exact result. \citet{Mash07} displayed the secular precessions of the argument of pericentre $\omega$, to order $\mathcal{O}(e)$, and of the longitude of the ascending node $\Omega$, to order $\mathcal{O}(e^2)$. While \rfr{nodus} agrees with (A.25) by \citet{Mash07}, \rfr{perius} is in disagreement with the sum of (A.24) for $\omega$ and the additional pericentre precession due to $\bds A = -K_1\bds R$ by \citet{Mash07}, even in the limit $e\rightarrow 0$. The variation of the mean anomaly was not computed by \citet{Mash07}.

It is also interesting to work out the variations after one orbital revolution of the radial, transverse and normal components of the perturbation of the radius vector $\bds R$.
By using\footnote{Here the shifts in the Keplerian orbital elements come from  indefinite integration(s), so that  they are functions of the fast angular variable used which, in our case, is $E$.} \citep{Casotto}
{\footnotesize{\begin{eqnarray}
  \Delta R &=& \rp{r}{a}\Delta a-a \cos f\Delta e + a e (1-e^2)^{-1/2}\sin f\Delta \mathcal{M}, \\ \nonumber \\
  \Delta\Xi &=& a\sin f\left[1+\rp{r}{a(1-e^2)}\right]\Delta e + r(\Delta \omega+\cos I\Delta\Omega) +\rp{a^2}{r}\sqrt{1-e^2}\Delta \mathcal{M}, \\ \nonumber \\
  \Delta\Upsilon &=& r\left(\sin u\Delta I-\cos u\sin I\Delta\Omega \right),
\end{eqnarray}
}
}
it is possible to obtain\footnote{Here $\Delta\xi=\Delta\xi(2\pi)-\Delta\xi(0),\ \xi=R,\Xi,\Upsilon$.}
{\footnotesize{\begin{eqnarray}
  \Delta R &=& \rp{5\pi ae\sqrt{1-e^2}\Delta K\sin^2 I\sin 2\omega}{2 n^2}, \lb{raggi}\\ \nonumber \\
  \Delta \Xi &=& \sqrt{\rp{1+e}{1-e}}\rp{\pi a\left\{(2+3e)\left[3K_1+K_2+\Delta K\cos 2I\right]+10e\Delta K\sin^2 I\cos 2\omega\right\}}{2 n^2},\lb{trav} \\ \nonumber \\
  \Delta\Upsilon &=& \rp{\pi a (-1+e)\sqrt{1-e^2}\Delta K\sin 2I\cos\omega}{2n^2}\lb{norma}.
\end{eqnarray}
}
}

Concerning the perturbation of the velocity vector $\bds V$, its radial, transverse and normal components can be expressed, in general, as \citep{Casotto}
{\footnotesize{
\begin{eqnarray}
  \Delta V_{R} &=& -\rp{n\sin f}{\sqrt{1-e^2}}\left(\rp{e}{2}\Delta a+\rp{a^2}{r}\Delta e\right)-\rp{n a^2\sqrt{1-e^2}}{r}\left(\Delta\omega+\cos I\Delta\Omega\right)-\rp{n a^3}{r^2}\Delta \mathcal{M}, \\ \nonumber \\
  \Delta V_{\Xi} &=& -\rp{na\sqrt{1-e^2}}{2r}\Delta a+\rp{na(e+\cos f)}{(1-e^2)^{3/2}}\Delta e+\rp{nae\sin f}{\sqrt{1-e^2}}\left(\Delta\omega+\cos I\Delta\Omega\right), \\ \nonumber \\
  \Delta V_{\Upsilon} &=& \rp{na}{\sqrt{1-e^2}}\left[\left(\cos u+e\cos\omega\right)\Delta I + \left(\sin u + e\sin\omega\right)\sin I\Delta\Omega\right].
\end{eqnarray}
}
}
From them it is possible to obtain\footnote{Also in this case, it is to be intended $\Delta V_{j}=\Delta V_j(2\pi)-\Delta V_j(0),\ j=R,\Xi,\Upsilon$.}
{\footnotesize{
\begin{eqnarray}
  \Delta V_{R} &=& \rp{\pi a\left\{
  \left[-4-3e(1+2e-e^2)\right]\left[3K_1+K_2+\Delta K\cos 2I\right]  +10e(-1-2e+e^2)\Delta K\sin^2 I\cos 2\omega
  \right\}}{4(1-e)^2n}, \\ \nonumber \\
  \Delta V_{\Xi} &=& -\rp{5\pi a e\Delta K\sin^2 I\sin 2\omega}{2(1-e)n}, \\ \nonumber \\
  \Delta V_{\Upsilon} &=& \rp{\pi a(1+4 e^2)\Delta K
\cos I\sin 2\omega}{2(1-e)n}.
\end{eqnarray}
}
}
While the transverse and normal components experience only long-term harmonic variations, the radial one exhibits a secular variation as well.
\section{Confrontation with the latest observational determinations}\lb{tria}
\subsection{The precessions of the perihelia of the inner planets of the solar system}
Recently, \citet{Pit010} has analyzed more than 550000 planetary observations of several kinds covering the time interval $1913-2008$. She used the dynamical force models of the EPM2008 ephemerides by estimating about 260 parameters.
Among them, she also determined corrections $\Delta\dot\varpi$ to the standard Newtonian/Einsteinain perihelion precessions of all the planets of the solar system including Pluto as well; such corrections, by construction, account for any unmodelled/mismodelled dynamical effects, so that they can be used, in principle, to put constraints on $K_1$ and $K_2$. To this aim, we will use the inner planets, whose estimated perihelion corrections are listed in Table \ref{tavolazza}, because they are more accurate.
\begin{table}
\centering
\caption{Estimated corrections $\Delta\dot\varpi$, in mas cty$^{-1}$ (1 mas cty$^{-1}=1.5\times 10^{-18}$ s$^{-1}$), to the standard perihelion precessions with the EPM2008 ephemerides. The quoted errors are not the formal, statistical ones but are realistic. From Table 8 of \citet{Pit010}.}
\label{tavolazza}
\begin{tabular}{lllll}\hline
& Mercury & Venus & Earth & Mars\\
\hline
$\Delta\dot\varpi$ (mas cty$^{-1}$) & $-4\pm 5$  & $24\pm 33$  & $6\pm 7$ & $-7\pm 7$\\
\hline
\end{tabular}
\end{table}
In Table \ref{tavolacoff} we compute the values of the coefficients $\mathcal{F}$ and $\mathcal{G}$, in s, for the inner planets of the solar system.
\begin{table}
\centering
\caption{Numerical values, in s, of the coefficients $\mathcal{F}$ and $\mathcal{G}$ for the inner planets according to \rfr{effe} and \rfr{gi}. The semimajor axis $a$, entering the Keplerian mean motion $n$, the eccentricity $e$, the inclination $I$ and the argument of perihelion $\omega$ have been retrieved from Table A.2 of \citet{Mur}. Reference frame: ICRF/J2000. Coordinate system: Ecliptic and Mean Equinox of Reference Epoch. }
\label{tavolacoff}
\begin{tabular}{lllll}\hline
& Mercury & Venus & Earth & Mars\\
\hline
$\mathcal{F}$ (s) & $-1.76948\times 10^6$  & $-4.61929\times 10^6$  & $-7.53291\times 10^6$ & $-1.40874\times 10^7$\\
$\mathcal{G}$ (s)& $-6238.21$  & $-15397.9$  & $-6.99848\times 10^{-6}$ & $-20198.2$\\
\hline
\end{tabular}
\end{table}
By equating the predicted precession of \rfr{peridot}, with the figures of Table \ref{tavolacoff},  to the estimated corrections $\Delta\dot\varpi$  listed in Table \ref{tavolazza},
it is possible to write down six inhomogeneous linear systems of two equations in the two unknowns $K_1$ and $K_2$. It turns out that the Earth and Mars yield the tightest constraints on $K_1$ and $K_2$: they are
\begin{eqnarray}
  K_1 &=& (-1.2\pm 1.4)\times 10^{-24}\ {\rm s}^{-2}, \lb{grande1}\\  \nonumber \\
  K_2 &=&  (1.4\pm 1.5)\times 10^{-21}\ {\rm s}^{-2}. \lb{piccolo1}
\end{eqnarray}
The quoted uncertainties have been obtained by linearly propagating the errors in $\Delta\dot\varpi$ of Table \ref{tavolazza}.

In Table \ref{tavolina} we quote the values of the corrections $\Delta\dot\varpi$ estimated by \citet{Pit05} with older versions of the EPM ephemerides and less extended data sets; for Venus, Earth and Mars they are, for some reasons, more accurate by about one order of magnitude than the more recent results of Table \ref{tavolazza}.
Using the values of the corrections $\Delta\dot\varpi$ of Table \ref{tavolina}
\begin{table}
\centering
\caption{Corrections $\Delta\dot\varpi$, in milliarcsec cty$^{-1}$ (1 mas cty$^{-1}=1.5\times 10^{-18}$ s$^{-1}$), to the standard perihelion precessions estimated by E.V. Pitjeva with the EPM2005 (Mercury, Earth, Mars) and EPM2006 (Venus) ephemerides. The quoted errors are not the formal, statistical ones but are realistic. From Table 3 of \citet{Pit05} (Mercury, Earth, Mars) and Table 4 of \citet{Fienga} (Venus).}
\label{tavolina}
\begin{tabular}{lllll}\hline
& Mercury & Venus & Earth & Mars\\
\hline
$\Delta\dot\varpi$ (mas cty$^{-1}$) & $-3.6\pm 5.0$  & $-0.4\pm 0.5$  & $-0.2\pm 0.4$ & $0.1\pm 0.5$\\
\hline
\end{tabular}
\end{table}
allows to obtain more stringent  constraints for $K_1$ and $K_2$. It turns out that, in this case, Venus and the Earth yield the tightest ones which are
\begin{eqnarray}
  K_1 &=& (4\pm 8)\times 10^{-26}\ {\rm s}^{-2}, \lb{grande}\\  \nonumber \\
  K_2 &=&  (3\pm 7)\times 10^{-23}\ {\rm s}^{-2}\lb{piccolo}.
\end{eqnarray}
In both cases, it turns out that
$K_1, K_2$ and $\Delta K$ are statistically compatible with zero; anyway, note that
$K_2$ may be up to three orders magnitude larger than $K_1$.

It may be interesting to use such values for $K_1$ and $K_2$, derived from the inner planets, with Saturn in order to see if they yield a LTB perihelion precession for the ringed planet compatible with its anomalous perihelion precession
\begin{eqnarray}
  \Delta\dot\varpi^{(\rm Pit\ I)}_{\rm Sat} &=& -0.006\pm 0.002\ {\rm arcsec\ cty}^{-1}, \lb{sat1}\\ \nonumber \\
  \Delta\dot\varpi^{(\rm Pit\ II)}_{\rm Sat} &=& -0.010\pm 0.015\ {\rm arcsec\ cty}^{-1}, \lb{sat2}\\ \nonumber \\
  \Delta\dot\varpi_{\rm Sat}^{(\rm Fie)} &=&  -0.010\pm 0.008\ {\rm arcsec\ cty}^{-1} \lb{sat3}
\end{eqnarray}
recently estimated by processing also some years of normal points from the Cassini spacecraft with  both the\footnote{The value of \rfr{sat1} has been communicated by E.V. Pitjeva to the author in November 2008, and it has explicitly been reported by \citet{Fienga} in their Table 4.} EPM and the INPOP ephemerides \citep{Fienga,Pit010}; see Section \ref{Cassini} for an analysis involving a different quantity related to such interplanetary ranging data.
To this aim, we have, first, to compute
\begin{eqnarray}
  \mathcal{F}_{\rm Sat} &=&  -2.21546\times 10^8\ {\rm s}, \\ \nonumber \\
  \mathcal{G}_{\rm Sat} &=& -22134.71  \ {\rm s}.
\end{eqnarray}
Then, it turns out that
\eqi \rp{\Delta\dot\varpi_{\rm Sat}-K_1{\mathcal{F}}_{\rm Sat} -K_2 {\mathcal{G}}_{\rm Sat}}{\delta(\Delta\dot\varpi_{\rm Sat})+|\mathcal{F}_{\rm Sat}|\delta K_1+|\mathcal{G}_{\rm Sat}|\delta K_2}<1\eqf for all the values of $\Delta\dot\varpi_{\rm Sat}$ (\rfr{sat1}-\rfr{sat3}) and the previously obtained figures for $K_1$ and $K_2$ (\rfr{grande1}-\rfr{piccolo1} and \rfr{grande}-\rfr{piccolo}).

\subsection{The Earth-Saturn range}\lb{Cassini}
Actually, it is possible to infer lower upper bounds by using  different targets.
After the Cassini spacecraft started its \virg{grand tour} of the Saturnian system, it has been possible to drastically increase the accuracy of the orbit determination of the ringed planet through direct ranging to Cassini itself. Figure B-20 of \citet{DE421} shows the range residuals of Saturn from 2004 to 2006 constructed with the DE421 ephemerides from Cassini normal points; processing of extended data records of Cassini is currently ongoing. The range residuals of Figure B-20 \citep{DE421} are accurate at 10 m level. Also a pair of range residuals from close encounters with Voyager 1 (1980) and Voyager 2 (1982) are shown: they are almost one order of magnitude less accurate. Thus, we will, now, look at the perturbation induced by $K_1$ and $K_2$ on the Earth-Saturn range $\bds \rho$ and we will compare it to the DE-421-based residuals, which are a somewhat more direct kind of observable with respect to the corrections to the perihelion precessions\footnote{The longitude of the perihelion, as all the other Keplerian orbital elements, are not directly observable quantities.}.
Figure \ref{rangesat} shows the numerically computed LTB Earth-Saturn range perturbation for the largest  values of \rfr{grande}-\rfr{piccolo} over a time interval of 2 yr, comparable to that of the available Cassini residuals.
\begin{figure}[ht!]
\includegraphics[width=\columnwidth]{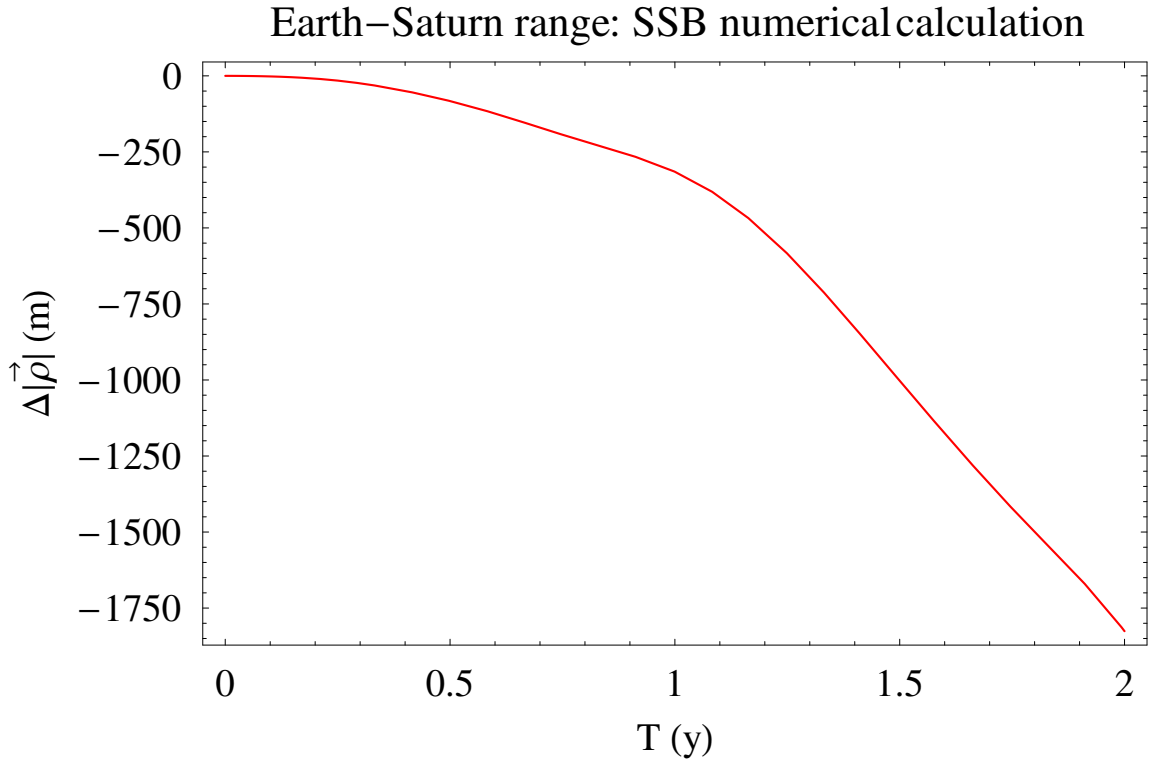}
 \caption{Difference $\Delta |\bds{\rho}|$ in the numerically integrated (Earth-Moon-Barycenter)-Saturn ranges with and without the nominal perturbation due to the LTB cosmological tidal effects for $K_1=12\times 10^{-26}$ s$^{-2}$ and $K_2=7\times 10^{-23}$ s$^{-2}$ over $\Delta T=2$ yr. The same initial conditions (J2000) have been used for both the integrations. The state vectors at the reference epoch have been retrieved from the NASA JPL Horizons system. The integrations have been performed in the  ICRF/J2000.0 reference frame centered at the Solar System Barycenter (SSB). }\lb{rangesat}
\end{figure}
The amplitude of the signal is 1750 m, which is  two orders of magnitude larger than the Cassini residuals. Even by assuming that part of the putative LTB signal, not explicitly modeled in the dynamical force models used to construct the residuals, has been absorbed and removed in the estimation of the initial conditions, it seems reasonable to conclude that a non-negligible part of a so huge anomalous effect would have been left in the residuals and, thus, it would  have not escaped from detection. It can be shown that by posing $K_1= K_2\approx 10^{-26}$ s$^{-2}$ yields a LTB signal as large as 100 m over 2 yr, while for $K_1=K_2\approx 10^{-27}$ s$^{-2}$ the amplitude of the LTB range is just 10 m over 2 yr.
Thus, we conclude that Saturn poses $10^{-27}$ s$^{-2}$ as upper bounds for both $K_1$ and  $K_2$.
It is likely that when the analysis of more extended Cassini data sets will be completed, more stringent limits on $K_1$ and $K_2$ will be placed.
\subsection{The Oort cloud}
The Oort cloud \citep{Oo}, populated by a huge number of small bodies moving along very eccentric orbits highly inclined to the ecliptic, is supposed to exist in the remote peripheries of the solar system. Indeed, it is a spheroid with a semimajor axis of about 100 kau elongated towards the Galactic center and a semiminor axis of about 80 kau \citep{Weis}. The interaction of a nearby passing star with the Oort comet cloud can give rise to comet showers reaching the region of the major planets \citep{Hills,Weis}. The formation time scale of such a comet shower is about 1 Myr. For example, according to \citet{Boby}, the star HIP 89 825 (GL 710) has a probability of 0.86 of penetrating the Oort cloud in the next $1.45\pm 0.06$ Myr. Moreover, its has  a smaller, non-zero probability ($1\times 10^{-4}$) of falling into the region $r< 1$ kau, thus potentially influencing the dynamics of the Edgeworth-Kuiper belt objects as well.

It is interesting to see the extent to which the LTB cosmological tidal effects may alter the standard picture of the Oort cloud.  To this aim, we will numerically integrate over a (Keplerian) orbital period the paths of a typical Oort comet\footnote{It can be shown that, in the Newtonian case, such initial conditions yield $X_{\rm max}=90$ kau, $Y_{\rm max}=60$ kau, $Z_{\rm max}=80$ kau over one Keplerian orbital period, so that it is well within  the domain indicated by \citet{Weis}.} ($X_0=40$ kau, $Y_0=30$ kau, $Z_0=5$ kau, $\dot X_0=-23$ kau Myr$^{-1}$, $\dot Y_0=-15$ kau Myr$^{-1}$, $\dot Z_0=-15$ kau Myr$^{-1}$ corresponding to  $a=66.7$ kau, $e=0.92$, $I=81.6$ deg), sharing the same initial conditions,  with and without the LTB tidal acceleration of \rfr{LTBacce}. Although not in a strict quantitative way as in the previous cases examined so far, such an approach may indicatively provide us with  tighter constraints on $K_1$ and $K_2$ by excluding those values  yielding too bizarre or implausible scenarios for the orbits of the Oort comets.

\begin{figure}
\centerline{
\vbox{
\epsfysize= 5.5 cm\epsfbox{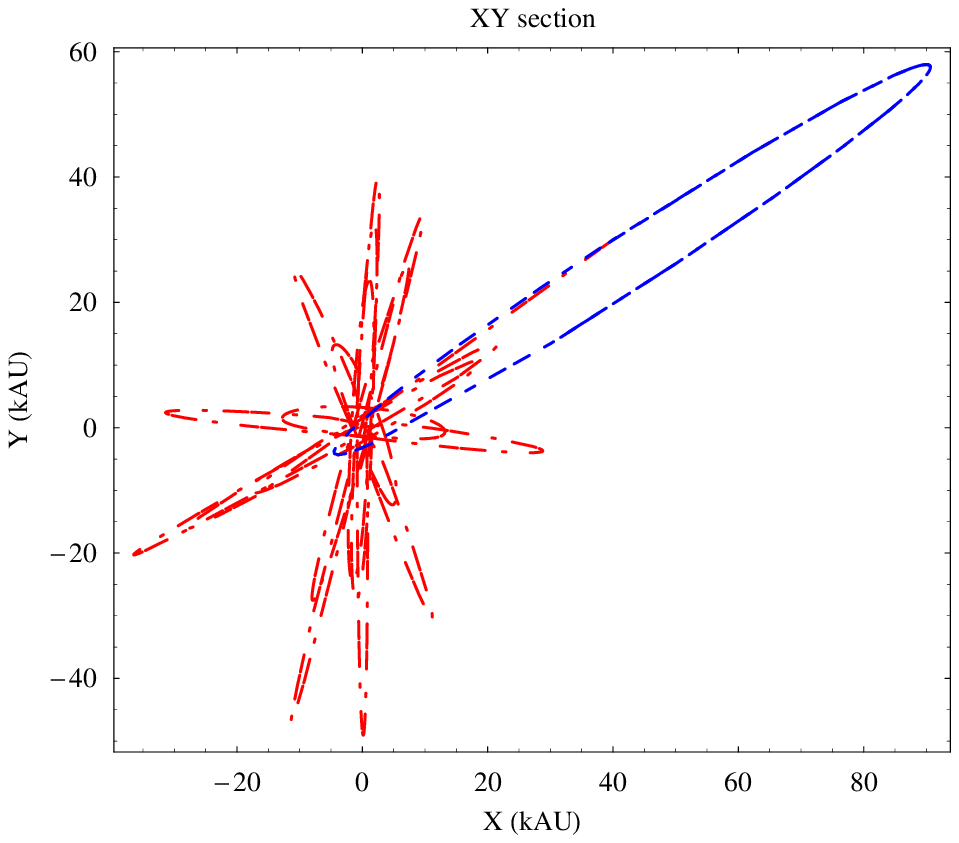}
\epsfysize= 5.5 cm\epsfbox{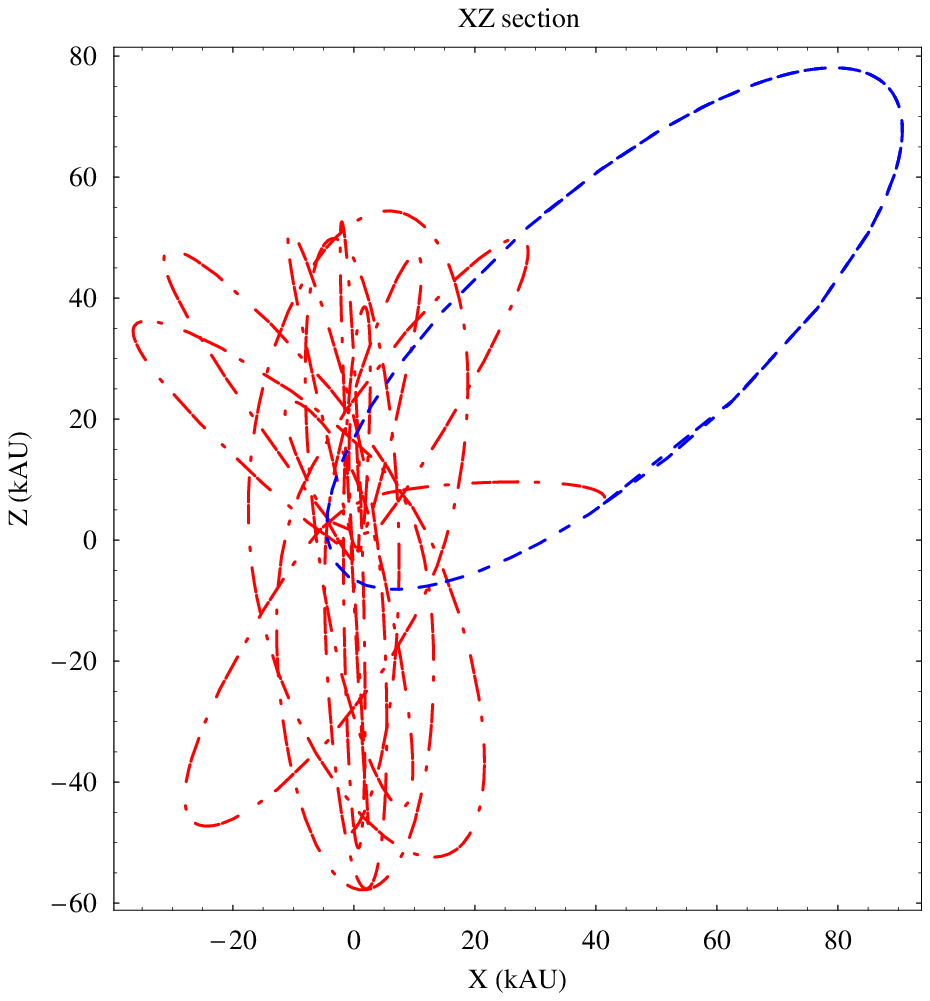}
\epsfysize= 5.5 cm\epsfbox{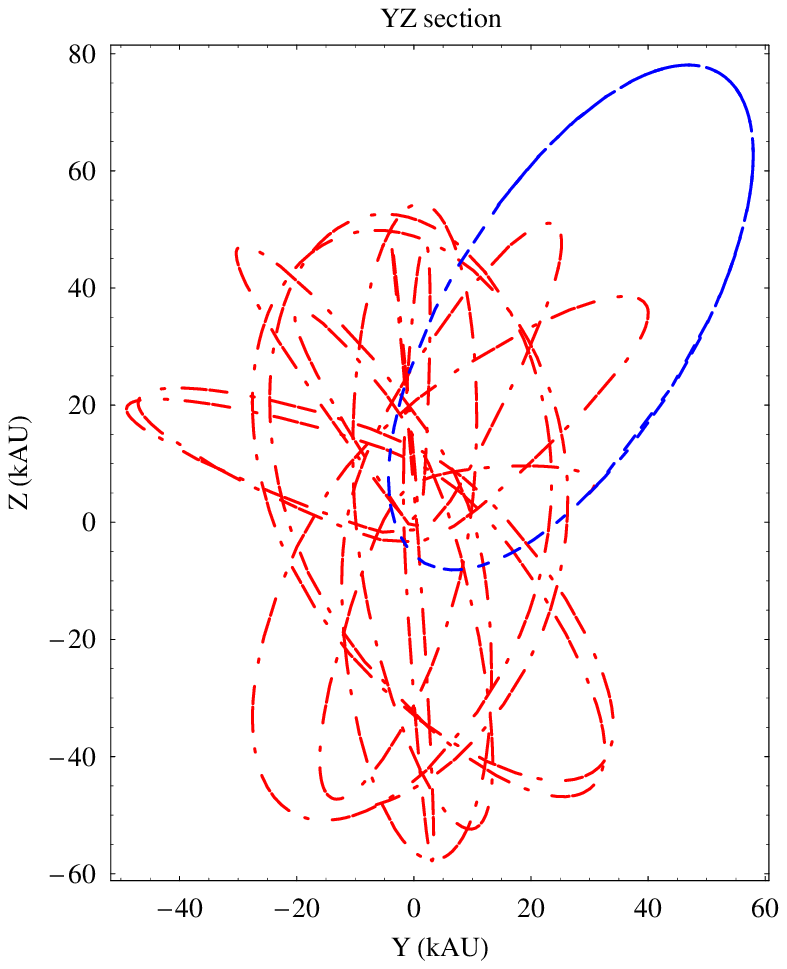}
      }
%
%
%
}
\caption{{Sections in the coordinate planes  of the numerically integrated orbits of an Oort comet with $a=66.7$ kau, $e=0.92$, $I=81.6$ deg. Dashed blue line: Newton. Dash-dotted red line: Newton+LTB with $K_1=4\times 10^{-27}$ s$^{-2}$, $K_2=3\times 10^{-27}$ s$^{-2}$. The initial conditions are $X_0=40\ {\rm kau}, Y_0=30\ {\rm kau}, Z_0=5\ {\rm kau},\dot X_0=-23\ {\rm kau\ Myr^{-1}},\dot Y_0=-15\ {\rm kau\ Myr^{-1}},\dot Z_0=-15\ {\rm kau\ Myr^{-1}}$ (1 kau Myr$^{-1}=0.0047$ km s$^{-1}$). The time span of the integration is $-P_{\rm b}\leq T\leq 0$.}\label{tripla}}
\end{figure}

\begin{figure}
\centerline{
\vbox{
\epsfysize= 5.5 cm\epsfbox{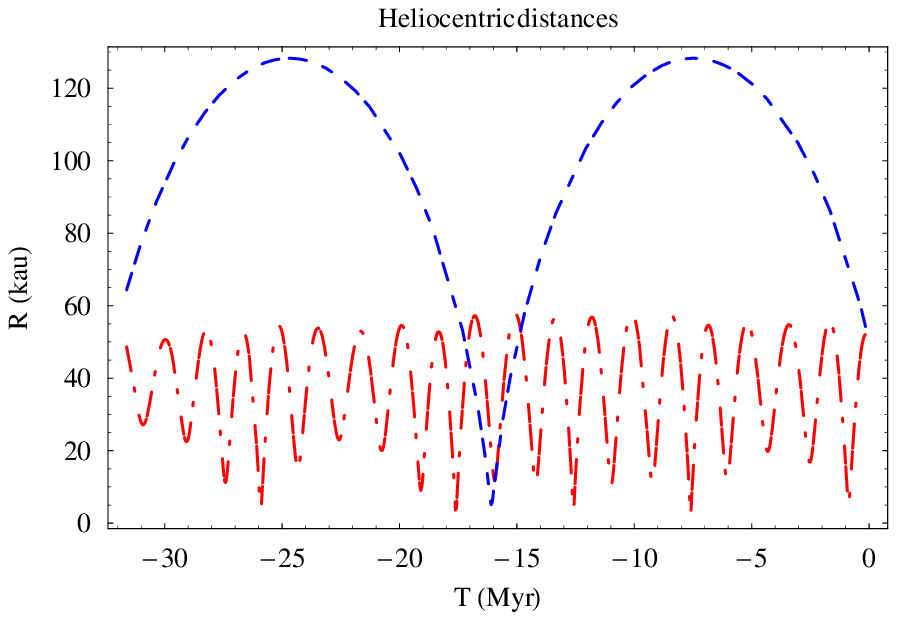}
\epsfysize= 5.5 cm\epsfbox{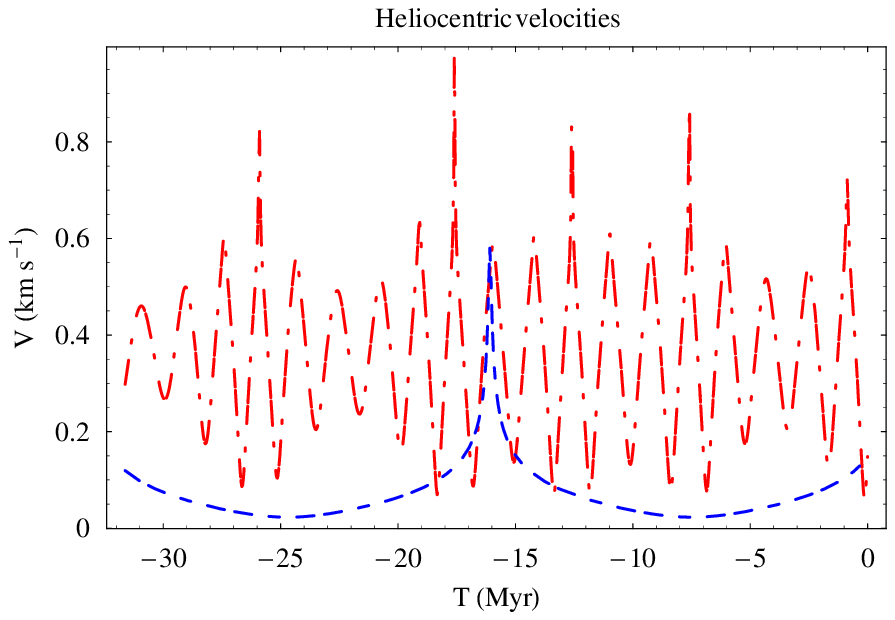}
      }
%
%
%
}
\caption{{Heliocentric distances, in kau, and velocities, in km s$^{-1}$,  for the numerically integrated orbits of an Oort comet with $a=66.7$ kau, $e=0.92$, $I=81.6$ deg. Dashed blue line: Newton. Dash-dotted red line: Newton+LTB with $K_1=4\times 10^{-27}$ s$^{-2}$, $K_2=3\times 10^{-27}$ s$^{-2}$. The initial conditions are $X_0=40\ {\rm kau}, Y_0=30\ {\rm kau}, Z_0=5\ {\rm kau},\dot X_0=-23\ {\rm kau\ Myr^{-1}},\dot Y_0=-15\ {\rm kau\ Myr^{-1}},\dot Z_0=-15\ {\rm kau\ Myr^{-1}}$ (1 kau Myr$^{-1}=0.0047$ km s$^{-1}$). The time span of the integration is $-P_{\rm b}\leq T\leq 0$.}\label{tripla_due}}
\end{figure}

Actually, Figure \ref{tripla}, depicting the sections in the coordinate planes of the purely Newtonian and LTB orbits for $K_1\approx K_2\approx 10^{-27}$ s$^{-2}$,  and Figure \ref{tripla_due}, showing the heliocentric distances and velocities  of the same paths, tell us that the values for the LTB parameters $K_1$ and $K_2$ allowed by the Earth-Saturn range are likely still too large. Indeed, the resulting numerically integrated paths are radically different from the Newtonian ones.
It must also be noted that the LTB acceleration ($3.0\times 10^{-11}$ m s$^{-2}$) is larger than the Newtonian one ($2\times 10^{-12}$ m s$^{-2}$), so that the perturbative scheme would not be applicable: the numerical integration of the equations of motion including both the LTB and the Newtonian accelerations is, thus, strictly required.
The consequences on the dynamics of the Oort cloud and on the comet showers may be not negligible, although, of course, not easily predictable. Suffices it to note from Figure \ref{tripla_due} the globally smaller spatial extension of the LTB orbit and its higher velocity which might reduce the impact of nearby passing stars\footnote{Clearly, also the motion of a perturbing star should be worked out in the framework of the LTB dynamics, but this is beyond the scope of the present paper.}.
Further numerical analyses show that reasonable orbits, i.e. not too far from the Newtonian picture, occur for $K_{1/2}\lesssim 10^{-30}-10^{-32}$ s$^{-2}$, when $A_{\rm LTB}/A_{\rm Newton}\approx 10^{-2}-10^{-4}$.

It must be stressed that it turns out that, actually, the LTB trajectories exhibit a strong dependence on the initial conditions. Thus, a complete statistical analysis involving vast ensembles of different initial conditions would be required, but it is outside the scope of the present work.
\section{Summary and conclusions}\lb{quatra}
We, first, analytically worked out the long-term, i.e. averaged over one orbital revolution, perturbations on the orbit of a  test particle moving in a local Fermi frame induced by the cosmological tidal effects of the inhomogeneous Lema\^{\i}tre-Tolman-Bondi model which has recently attracted attention as possible explanation of the observed cosmic acceleration without resorting to dark energy. In particular, we computed the variations of the semimajor axis $a$, the eccentricity $e$, the inclination $I$, the longitude of the ascending node $\Omega$, the argumet of pericenter $\omega$ and the mean anomaly $\mathcal{M}$ by means of the Lagrange's variational equations with the eccentric anomaly $E$ as fast variable. We found that $a$ does not experience any change, on average, while the other Keplerian orbital elements undergo long-term variations including both secular and harmonic terms with frequency $2\omega$; $e$ and $I$ exhibit only sinusoidal changes, while $I,\Omega,\mathcal{M}$  secularly precess as well.
We repeated the calculation also with the Gauss perturbative approach  by finding the same results.
We also explicitly computed the changes  over one orbital revolution of the position and velocity vectors along the radial $\bds{\hat R}$, transverse $\bds{\hat \Xi}$ and normal directions $\bds{\hat \Upsilon}$ of a frame co-moving with the test particle. While the radial and normal components  $\Delta R$ and $\Delta\Upsilon$ of the perturbation of the position vector experience only harmonic variations, the transverse one $\Delta\Xi$ shows secular changes as well. Concerning the velocity, its radial perturbation $\Delta  V_R$ undergoes both secular and sinusoidal modifications, while its transverse and normal components $\Delta V_{\Xi}$ and $\Delta V_{\Upsilon}$ exhibit harmonic signatures only. In obtaining our results we  made no approximations on $e$ and $I$ by retaining all terms.

Then, we phenomenologically put constraints on both the parameters $K_1$ and $K_2$ of the LTB metric in the local Fermi frame  by looking at various astronomical bodies and data of the solar system. By comparing our analytical prediction for the rate of the longitude of pericenter $\varpi$ to different sets of the corrections $\Delta\dot\varpi$ to the standard Newtonian/Einsteinian precessions of the perihelia of the inner planets recently estimated with the EPM ephemerides we found that the tightest constraints are $K_1 = (4\pm 8)\times 10^{-26}$ s$^{-2}$, $K_2 = (3\pm 7)\times 10^{-23}$ s$^{-2}$. The confrontation of the residuals of the Earth-Saturn range, obtained by processing some years of radiotechnical data from the Cassini spacecraft as well, with the numerically computed LTB-induced Earth-Saturn range signal allowed to set  $K_1\approx K_2\approx 10^{-27}$ s$^{-2}$. By looking at the LTB-type distortions of the orbit of a typical object of the Oort cloud with respect to the commonly accepted Newtonian picture, based on the observations of the comet showers from that remote region of the solar system, pointed towards even lower bounds, i.e. $K_1\approx K_2\lesssim 10^{-30}-10^{-32}$ s$^{-2}$. According to cosmological data,  
$K_1\approx K_2=-4\times 10^{-36}$ s$^{-2}$.

\end{document}